\documentclass[pra,twocolumn,showpacs,superscriptaddress,groupedaddress]{revtex4-1}
\usepackage{amsfonts}
\usepackage{amsmath}
\usepackage{amssymb}
\usepackage{graphicx}

\input{tcilatex}
\begin{document}

\title{Equilibrium vortex configurations in ultra-rapidly rotating
two-component Bose-Einstein condensates}
\author{C.-H. Hsueh$^{1}$, T.-L. Horng$^{2}$, S.-C. Gou$^{3}$, and W. C. Wu$%
^{1}$}
\date{\today}

\begin{abstract}
The equilibrium vortex formations in rotating binary Bose gases with a
rotating frequency higher than the harmonic trapping frequency is
investigated theoretically. We consider the system being evaporatively
cooled to form condensates and a combined numerical scheme is applied to
ensure the binary system being in an authentic equilibrium state. To keep
the system stable against the large centrifugal force of ultrafast rotation,
a quartic trapping potential is added to the existing harmonic part. Using
the Thomas-Fermi approximation, a critical rotating frequency $\Omega_{c}$
is derived, which characterizes the structure with or without a central
density hole. Vortex structures are studied in details with rotation
frequency both above and below $\Omega_{c}$, and with respect to the
miscible, symmetrically-separated, and asymmetrically-separated phases in
their non-rotating ground-state counterparts.
\end{abstract}

\maketitle

\affiliation{$^1$Department of Physics, National Taiwan Normal University, Taipei 11677, Taiwan\\
$^2$Department of Applied Mathematics, Feng Chia University, Taichung 40724, Taiwan\\
$^3$Department of Physics, National Changhua University of Education, Changhua
50058, Taiwan}

\section{Introduction}

Quantum coherence has enabled intriguing phenomenon, such as quantized
vorticity, in Bose-Einstein condensates (BECs). When a trapped condensate is
driven to rotate, singly quantized vortices form. In lower rotations, only
one or few vortices will be present at equilibrium \cite{vortex1}. Faster
rotation can generate more vortices which are eventually condensed into a
lattice \cite{lattice1,lattice2,lattice3,lattice4}. Vorticity in a
single-component Bose condensates has indeed been observed in a variety of
experiments. On the other hand, since the first experiment of two coexisting
condensates of two different hyperfine states of $^{87}$Rb \cite{exp} was
realized, BEC in mixtures of trapped quantum gases provides a unique
opportunity to study the miscibility of interpenetrating quantum fluid.
Several theoretical articles about binary-mixture condensates have expounded
that both the interspecies and intraspecies interactions play an important
role in determining the density patterns and phase separation of the
condensates \cite{sym-asym1,sym-asym,sym-asym2,sym-asym3,sym-asym4,sym-asym5}%
. In contrast to the Abrikosov vortex-lattice state of a scalar BEC, the
vortex states of binary-mixture BECs have various exotic structures due to
the variety of interactions \cite{rotate two,rotate two1,rotate two2}.

Rotating Bose condensates are usually confined in a harmonic trap with
cylindrical symmetry around the rotation axis (say, $z$-direction). In these
typical cases, there are two limiting regimes depending on the relative size
of the rotating frequency $\Omega_{0}$ and the trapping frequency $\omega$
in the $xy$ plane. When $\Omega_{0}>\omega$, the system will become unstable
due to a strong centrifugal force. In order to analyze the regime of
ultrafast rotations with $\Omega_{0}>\omega$, one approach is to add a
quartic part to the harmonic potential. In this type of system, the trapping
force will be always greater than the centrifugal force and consequently the
regime $\Omega_{0}>\omega$ can be fully explored \cite{multiply
quantized1,multiply quantized2,multiply quantized3,multiply
quantized4,multiply quantized5,multiply quantized6,multiply
quantized7,multiply quantized8}. The current paper attempts to study the
equilibrium vortex states of ultrafast-rotating binary condensates confined
in a harmonic-plus-quartic potential.

In a single-component fast-rotating condensate, when $\Omega_{0}>\omega$,
the system can experience an effective potential of Mexican-hat shape.
Depending on how $\Omega_{0}$ is larger than $\omega$, the system can be
roughly separated into two regimes: a condensate with or without the central
density hole. More exactly, it has been shown in the literature that there
exist three distinct phases for the fast-rotating scalar condensate confined
in a harmonic-plus-quartic potential. One is the vortex lattice without a
hole (VL), the second is the vortex lattice with a hole (VLH), and the third
is the giant vortex state (GV) \cite{multiply quantized4,multiply quantized7}%
. It is interesting to see how the interspecies and intraspecies
interactions play the role in the binary-mixture condensates under fast
rotation and in particular, how the above-mentioned phases manifest in these
systems.

Due to the complexity of the interactions in the binary system, a standard
imaginary-time propagating method may not be easy to find the converging
results for the true equilibrium states. It has been shown in Ref.~\cite%
{phenomenological damping} that Gross-Pitaevskii equation (GPE) with a
phenomenological damping term can provide an efficient numerical machinery
for finding the eigenstates of the time-independent GPE. This, \textbf{a
similar} stochastic Gross-Pitaevskii equation (SGPE) approach \cite{SGPE
Stoof1,SGPE Stoof2,SPGPE Gardiner1,SPGPE Gardiner2,vortex SPGPE}, has been
demonstrated to be an efficient way for studying the single-component BEC
system. It is also anticipated that SGPE is an alternatively efficient
method for studying the dynamic and equilibrium properties of a
multi-component system near absolute zero.

The paper is organized as follows. In Sec.~\ref{sec:methodology}, we
introduce the theory for studying the equilibrium vortex states of a
binary-mixture BEC system. To investigate the regime of ultrafast rotation,
the system is trapped under a harmonic-plus-quartic potential. 
Sec.~\ref{sec:results} is devoted to a detailed discussion of the vortex
structures for fast-rotating binary-mixture BECs. A critical rotation
frequency $\Omega _{c}$ is derived and both $\Omega <\Omega _{c}$ and $%
\Omega >\Omega _{c}$ regimes are studied. It will be shown explicitly that
vortex structures of the system do manifest the ground states of their
non-rotating counterparts. Sec.~\ref{sec:conclusion} is a brief conclusion.

\section{Methodology}

\label{sec:methodology}

We consider rapidly rotating two-component pancake-shape BECs that are
parallel to the $xy$-plane and in a cylindrically symmetric potential.
Assuming that the excitation in the $z$-direction is suppressed, the system
can be treated approximately by a two-dimensional theory. In the mean-field
approximation, such a two-component BEC system in a co-rotating frame with a
rotating frequency $\Omega _{0}$ around the $z$-axis can be described by the
time-dependent coupled GPEs: 
\begin{eqnarray}
i\hbar \frac{\partial \Psi _{j}}{\partial t} &=&\left( \mathcal{L}_{\mathrm{%
GP}}^{\left( j\right) }-\mu _{j}\right) \Psi _{j}  \notag \\
&=&\left( \mathcal{H}_{j}+\sum\limits_{k=1,2}U_{jk}\left\vert \Psi
_{k}\right\vert ^{2}-\mu _{j}\right) \Psi _{j},  \label{GPE}
\end{eqnarray}%
where for component $j$ ($=1,2$), $\mathcal{L}_{\mathrm{GP}}^{\left(
j\right) }$ is the GP Hamiltonian, $\mu _{j}$ is the chemical potential, and 
$\Psi _{j}$ is the macroscopic wave function normalized under $N_{j}=\int
|\Psi _{j}|^{2}dxdy$ with $N_{j}$ the particle number. $\mathcal{H}%
_{j}\equiv -\hbar ^{2}\nabla ^{2}/\left( 2m_{j}\right) +V_{j}-\Omega
_{0}L_{z}$ is the single-particle Hamiltonian with $m_{j}$ the atomic mass, $%
V_{j}=m_{j}\omega _{j}^{2}r^{2}/2+u_{j}r^{4}/4$ the trapping potential in 
\textbf{polar} coordinates $\left( r,\phi \right) $, $\omega _{j}$ the
harmonic trapping frequency, $u_{j}$ the strength of the quartic potential,
and $L_{z}=-i\hbar \partial /\partial \phi $ the $z$-component angular
momentum operator. The interaction parameter $U_{jk}=2\pi \hbar ^{2}\tilde{a}%
_{jk}\left( m_{j}^{-1}+m_{k}^{-1}\right) $ with $\tilde{a}_{jk}(>0)$ the
effective two-dimensional $s$-wave scattering length between atoms in
components $j$ and $k$.

As mentioned before, it has been shown that GPE with a phenomenological
damping term \cite{phenomenological damping} and SGPE may be an efficient
numerical approach to obtain accurate ground states of a given
time-independent GPE. Here we shall apply a similar SGPE approach to study
the vortex states of a ultrafast-rotating two-component BEC system. The
coupled SGPEs for the present binary system can be expressed as 
\begin{equation}
i\hbar \frac{\partial \Psi _{j}}{\partial t}=\left( 1-i\gamma \right) \left( 
\mathcal{L}_{\mathrm{GP}}^{\left( j\right) }-\mu _{j}\right) \Psi _{j}+\eta
_{\gamma },  \label{SGPE}
\end{equation}%
where $\eta _{\gamma }=\eta _{\gamma }(\mathbf{r},t)$ is a complex Gaussian
noise considered arising due to the contact with the thermal modes. The
correlation function associated with the noise is given by $\left\langle
\eta _{\gamma }^{\ast }\left( \mathbf{r},t\right) \eta _{\gamma }\left( 
\mathbf{r}^{\prime },t^{\prime }\right) \right\rangle =2\hbar k_{B}T\gamma
\delta \left( \mathbf{r}-\mathbf{r}^{\prime }\right) \delta \left(
t-t^{\prime }\right) $. The strength of the noise, and hence the damping, is
thus proportional to $\gamma $. In principle, $\gamma $ can be calculated $ab
$ $initio$ in terms of the Keldysh self-energy \cite{SGPE Stoof1,SGPE Stoof2}%
. However, as we are interested only in the properties at equilibrium, we
may approximate it as a spatial and temporal constant. Throughout this
paper, $\gamma $ is reasonably taken to be a small number, $0.01$. The
temperature is set to be 1nK which is about $10^{-2}$ or less of the
critical temperature of a typical BEC system. Here we have to state that the
coupled SGPEs (\ref{SGPE}) do not describe the dynamics of a two-component
system accurately since any coupling of the thermal cloud components of the
two different condensates is not considered. Because we focus on the
equilibrium states, the effect of these couplings should be slight, and Eq. (%
\ref{SGPE}) is good enough for obtaining the equilibrium states of a
two-component system.

To reduce the number of parameters, we shall assume that $m_{1}=m_{2}\equiv
m $, $\omega _{1}=\omega _{2}\equiv \omega ,$ $u_{1}=u_{2}\equiv u$, and $%
N_{1}=N_{2}\equiv N$, respectively. Moreover, for convenience, the
computations will be carried in the oscillator units. That is, the length,
time, and energy are scaled respectively in units of $\sqrt{\hbar /m\omega }$%
, $1/\omega $, and $\hbar \omega $. As a consequence, the coupled SGPEs (\ref%
{SGPE}) take the following dimensionless forms: 
\begin{eqnarray}
&&i\frac{\partial \psi _{j}}{\partial t}=\left( 1-i\gamma \right) \times 
\notag \\
&&\left( \mathcal{-}\frac{\nabla ^{2}}{2}+\frac{r^{2}}{2}+\frac{\lambda r^{4}%
}{4}+i\Omega \frac{\partial }{\partial \phi }+\sum\limits_{k=1,2}g_{jk}\left%
\vert \psi _{k}\right\vert ^{2}-\tilde{\mu}_{j}\right) \psi _{j}  \notag \\
&&~~~~~+\tilde{\eta}_{\gamma }~.  \label{dimensionless}
\end{eqnarray}%
Here we have redefined the normalized wave function $\psi _{j}\equiv \sqrt{%
\hbar /\left( m\omega N\right) }\Psi _{j}$, the strength of the quartic trap 
$\lambda \equiv u\hbar /\left( m^{2}\omega ^{3}\right) $, the interaction
constants between atoms $g_{jk}\equiv 4\pi N\tilde{a}_{jk}$, the chemical
potential $\tilde{\mu}_{j}\equiv \mu _{j}/\hbar \omega $, and the noise $%
\tilde{\eta}_{\gamma }\equiv {\eta }_{\gamma }/\hbar \omega $. Besides the
rotation rate $\Omega \equiv \Omega _{0}/\omega $. The rotation rate and the
quartic trap strength will be fixed at $\Omega =2.5$ and $\lambda =1$ in our
calculation throughout this paper.

\begin{figure}[t]
\begin{center}
\includegraphics[width=8cm]{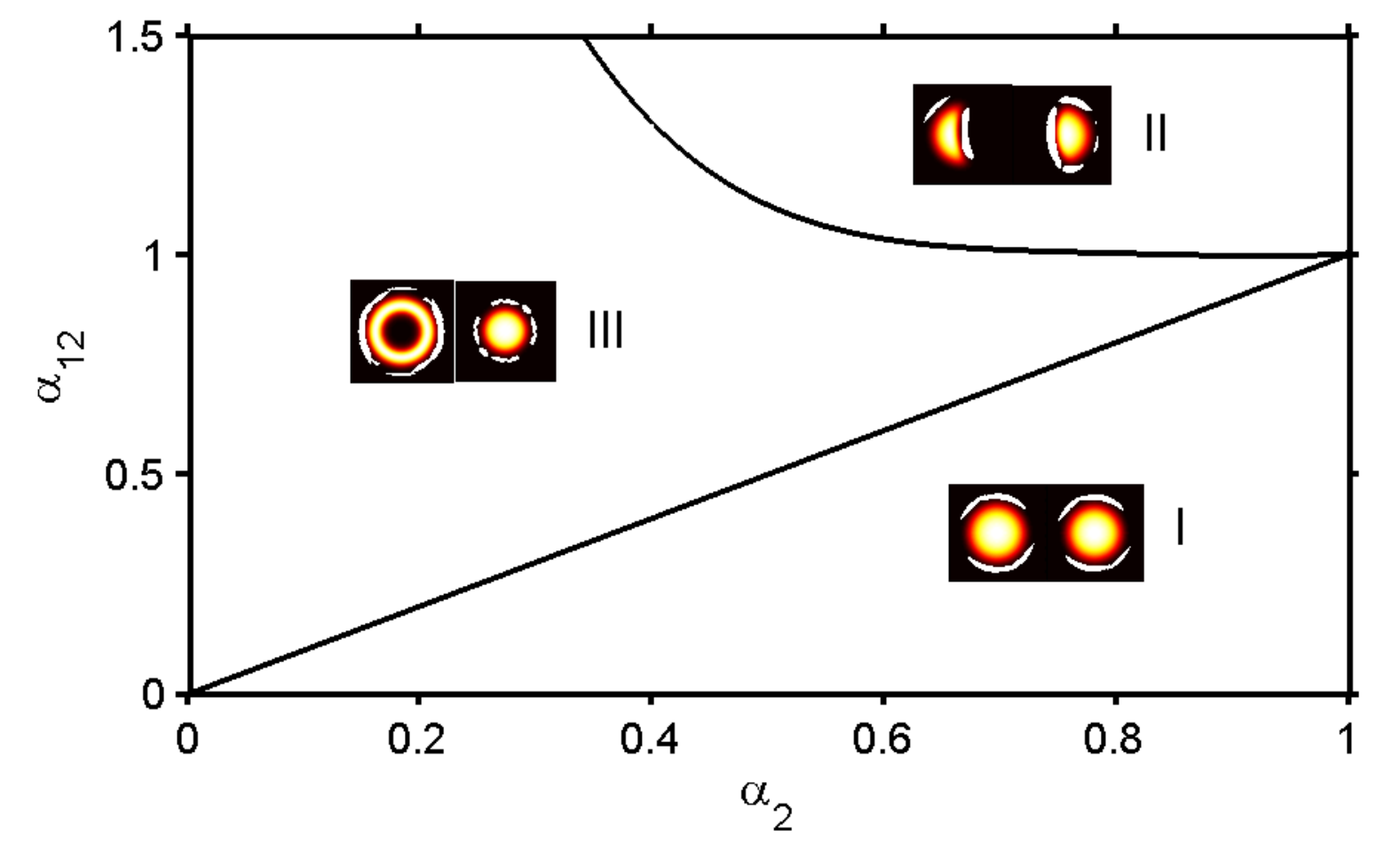}
\end{center}
\par
\vspace{-0.5cm}
\caption{Phase diagram of the non-rotating binary-mixture condensates
confined in a harmonic-plus-quartic trap in terms of the relative
interaction strengths, $\protect\alpha_{2}\equiv\tilde{a}_{22}/\tilde{a}%
_{11} $ and $\protect\alpha_{12}\equiv\tilde{a}_{12}/\tilde{a}_{11}$. The
quartic trap strength is fixed at $\protect\lambda=1$. Regions I, II, and
III correspond to miscible, asymmetrically separated, and symmetrically
separated phases, respectively.}
\label{fig1}
\end{figure}

In our calculations, the equilibrium solutions and the corresponding
chemical potentials are obtained by solving the norm-preserving
imaginary-time propagation of the time-dependent coupled GPEs (\ref{GPE})
starting from an arbitrary trial wave function. The propagation continues
until the fluctuation in the norm of the wave function becomes smaller than $%
10^{-5}$. To determine whether the vortex states obtained are indeed at
equilibrium, the solutions of the GPEs, which were converged by the
imaginary-time method, are then substituted into and treated as the initial
states of the coupled SGPEs (\ref{SGPE}). If the initial state was not an
equilibrium state, it would keep propagating until the damping term
vanishing.

Moreover, we have used the method of lines with spatial discretization by
the Fourier pseudospectral method to compute Eqs.~(\ref{GPE}) and (\ref{SGPE}%
) [or (\ref{dimensionless})]. The time integration in Eq.~(\ref{GPE}) is
done by the adaptive Runge-Kutta method of order 2 and 3 (RK23), which is
more time efficient due to an adjustable time step. However, the
fourth-order Runge-Kutta method (RK4) is used for Eq.~(\ref{SGPE}) [or (\ref%
{dimensionless})], since the thermal noise term $\eta _{\gamma}$ is
proportional to $1/\sqrt{dt}$, and is better computed with a fixed time step.

\section{Results and discussions}

\label{sec:results}

\subsection{Non-rotating ground states}

Solving the non-rotating ($\Omega =0$) time-dependent coupled GPEs (\ref{GPE}%
) using the imaginary-time propagating method, we have obtained three
distinct phases for the ground states of the binary-mixture condensates.
Fig.~\ref{fig1} shows the phase diagram of the binary-mixture condensates in
terms of the relative interaction strengths, $\alpha _{2}\equiv \tilde{a}%
_{22}/\tilde{a}_{11}$ and $\alpha _{12}\equiv \tilde{a}_{12}/\tilde{a}_{11}$%
. By symmetry, it is sufficient to consider $\alpha _{2}\leq 1$ only. The
non-rotating binary-mixture condensates are phase miscible in region I and
phase separated in regions II (asymmetric) and III (symmetric). As shown in
Fig.~\ref{fig1}, the boundary between phases I and III is linear, $\alpha
_{12}=\alpha _{2}$, which is obtained by jointing nine points: from $\alpha
_{2}=0.2$\ to $\alpha _{2}=1$\ spaced by $0.1$, and theirs corresponding $%
\alpha _{12}$\ are determined numerically with precision less than $0.01$.
The quartic trap ($\lambda $) has only a minimal effect on the boundary
between the miscible and the symmetric separated phases. On the contrary,
the boundary between phases II and III is quite $\lambda $-dependent, which
is obtained by jointing nine points: from $\alpha _{2}=0.4$\ to $\alpha
_{2}=1$\ spaced by $0.1$, and theirs corresponding $\alpha _{12}$\ are
determined numerically with precision less than $0.01$. Due to a relatively
stronger confinement, the area of phase II can expand as $\lambda $
increases. This allows a larger space for studying the asymmetric
phase-separated regime for fast-rotating binary-mixture condensates. The two
boundaries intersect at $\left( \alpha _{2},\alpha _{12}\right) =\left(
1,1\right) $ which is called the \emph{isotropic} point. We shall study the
vortex structures in all three phases and in particular at the isotropic
point.

\begin{figure}[t]
\begin{center}
\vspace{-0.5cm} \includegraphics[width=7.5cm]{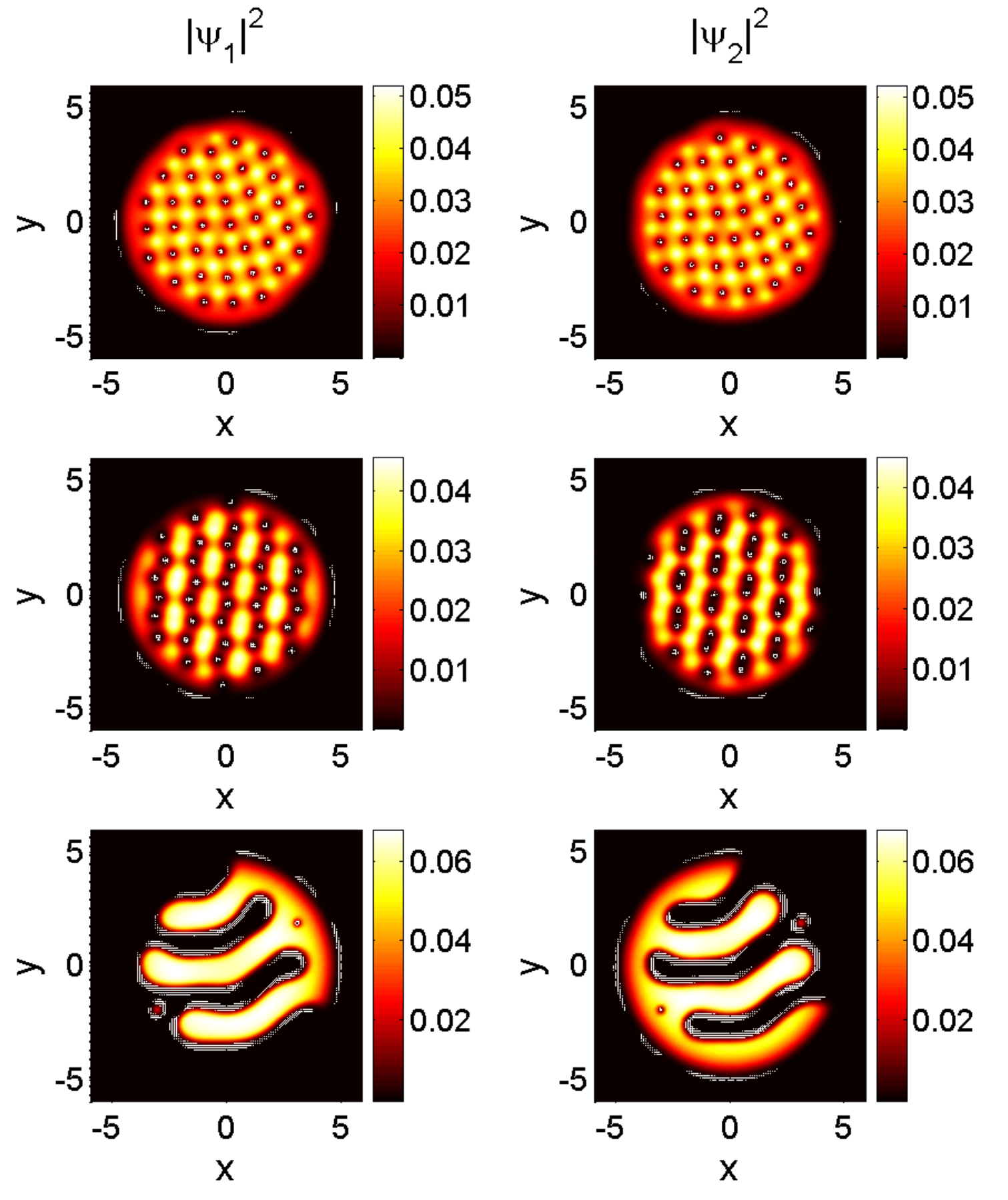}
\end{center}
\par
\vspace{-0.2cm}
\caption{(Color online) Vortex structures of individual component in
fast-rotating binary-mixture condensates confined in a harmonic-plus-quartic
potential. The interaction constants $g_{11}=g_{22}=1300$ ($\protect\alpha %
_{2}=1$), the rotation rate $\Omega =2.5$, and the quartic trap strength $%
\protect\lambda =1$ for all frames. From the top to the bottom rows, $%
\protect\alpha _{12}=0.7$, $1$, and $1.3$, respectively.}
\label{fig2}
\end{figure}

\subsection{Vortex states}

Before discussing the fast-rotating binary-mixture condensates, it is useful
to first examine the case of a scalar rotating condensate trapped in a
harmonic-plus-quartic potential. In the one-component system, there is only
one interaction constant $g\equiv 4\pi N\tilde{a}$, and when $g\gg 1$ the
system is in the so-called Thomas-Fermi (TF) regime. The TF density can be
obtained by ignoring the kinetic energy:%
\begin{equation}
n_{\mathrm{TF}}\equiv \left\vert \psi _{\mathrm{TF}}\right\vert ^{2}=\left[
\mu _{\mathrm{TF}}+\frac{\left( \Omega ^{2}-1\right) r^{2}}{2}-\frac{\lambda
r^{4}}{4}\right] /g.  \label{TF density}
\end{equation}%
For $\Omega <1$, the density has a local maxium near the center, but it
changes to a local minium for $\Omega >1$. Eq. (\ref{TF density}) has
solutions where the TF density vanishes%
\begin{equation}
R_{\gtrless }^{2}=\frac{(\Omega ^{2}-1)\pm \sqrt{4\lambda \mu _{\mathrm{TF}%
}+(\Omega ^{2}-1)^{2}}}{\lambda }.  \label{TF radii}
\end{equation}%
Here the upper (plus) sign denotes the outer radius $R_{>}$\ for any value
of the chemical potential $\mu _{\mathrm{TF}}$. In contrast, the lower
(minus) sign yields a physical inner radius $R_{<}$\ only if $\mu _{\mathrm{%
TF}}$\ is negative. Correspondingly the negative chemical potential can be
fixed by $\mu _{\mathrm{TF}}=(3g\sqrt{\lambda }/\pi )^{2/3}/4-(\Omega
^{2}-1)^{2}/(4\lambda )$ and the spatial extension of the atomic cloud can
be characterized by $R_{>}$ and $R_{<}$ which correspond to the outer and
inner TF radii, respectively. Accordingly, the system can be separated into
two regimes demarcated by a critical rotating frequency $\Omega _{c}$
corresponding to $R_{<}(\Omega _{c})=0$ [or $\mu _{\mathrm{TF}}(\Omega
_{c})=0$]. It is found that $\Omega _{c}^{2}=1+(3\lambda ^{2}g/\pi )^{1/3}$
which is dependent of the interaction constant $g$ and the quartic trap
strength $\lambda $. At sufficiently high rotating frequencies such that $%
\Omega >\Omega _{c}$ and hence $R_{<}>0$, a central hole will appear in the
condensates, \textrm{i.e.}, in the VLH state \cite{multiply
quantized4,multiply quantized7}. In contrast for $\Omega <\Omega _{c}$,
vortex lattice without a central hole regime (the VL state) will appear.

For the present binary-mixture condensates, the critical rotating frequency $%
\Omega _{c}$ which characterizes the transition between the VL and the VLH
states, can be qualitatively determined by the larger of $g_{11}$ and $g_{22}
$. As mentioned before, in this paper we consider only the cases $g_{11}\geq
g_{22}$ (i.e., $\alpha _{2}\leq 1$). Consequently 
\begin{equation}
\Omega _{c}^{2}\equiv 1+\left( 3\lambda ^{2}g_{11}/\pi \right) ^{\frac{1}{3}%
}.  \label{omegac}
\end{equation}%
When $\Omega <\Omega _{c}$, the rotating binary system will be in the VL
state, while when $\Omega >\Omega _{c}$, the system will be in the VLH
state. In the following two subsections, two distinct cases of $g_{11}=1300$
and $g_{11}=55$ will be studied. The former corresponds to a critical
rotating frequency $\Omega _{c}=3.43$ and hence $\Omega =2.5<\Omega _{c}$,
while the latter corresponds to $\Omega _{c}=2.18$ and hence $\Omega >\Omega
_{c}$. In the following calculations, the equilibrium vortex solutions and
the corresponding chemical potentials are obtained by solving the
norm-preserving imaginary-time propagation of the time-dependent coupled
GPEs (\ref{GPE}) starting from an arbitrary trial wave function. To
determine whether the vortex states obtained are indeed at equilibrium, the
solutions of the GPEs, which were converged by the imaginary-time method,
are then substituted into and treated as the initial states of the coupled
SGPEs (\ref{SGPE}).

\begin{figure}[t]
\begin{center}
\vspace{-0.5cm} \includegraphics[width=8.0cm]{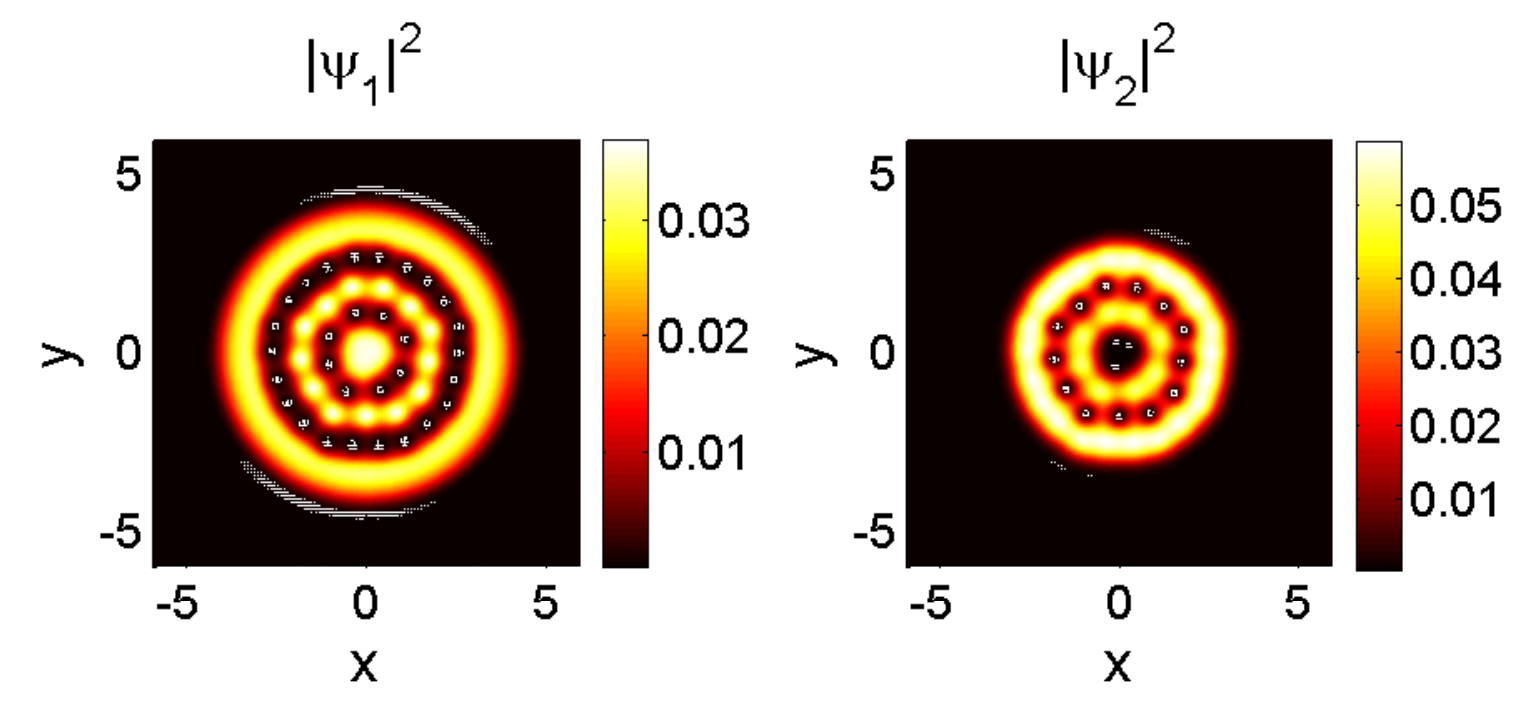}
\end{center}
\par
\vspace{-0.5cm}
\caption{(Color online) Vortex structures of component 1 (left panel) and 2
(right panel) in fast-rotating binary-mixture condensates. Here $g_{11}=1300$%
, $\protect\alpha _{2}=0.5$, $\protect\alpha _{12}=0.8$, $\protect\lambda =1$%
, and $\Omega =2.5$.}
\label{fig3}
\end{figure}

\subsubsection{VL state with $\Omega<\Omega_{c}$}

We first consider the vortex structures with rotation frequency below the
critical rotating frequency, $\Omega <\Omega _{c}$. Fig.~\ref{fig2} shows
the equilibrium vortex structures of two-component condensates confined in a
harmonic-plus-quartic potential with $g_{11}=g_{22}=1300$ ($\alpha _{2}=1$)
and $\alpha _{12}=0.5$, $1$, and $1.3$ (from the top to the bottom)
respectively. All three cases are belonging to the VL regime to which one is
able to conclude for the following. (i) For a phase-miscible mixture, the
equilibrium state is composed of regular vortex lattices which form roughly
a square lattice at $\alpha _{12}=0.5$ than what is expected to be a
triangular lattice when $\alpha _{12}\rightarrow 0$ \cite{antiferromagnetism}%
. (ii) At the isotropic point ($\alpha _{2}=\alpha _{12}=1$), a honeycomb
lattice is formed for one component, while vortices in the other component
form a vortex-pair lattice (vortex of every pair has the same circulation).
(iii) Stationary vortex sheets are formed for an asymmetric phase-separated
mixture. Our results in Fig.~\ref{fig2} are intended to be compared to those
shown in Figs.~2(a), 3, and 4 in Ref.~\cite{rotate two1}. With our results,
we have been able to verify that the one shown in Fig.~3(b) of Ref.~\cite%
{rotate two1} corresponds to an authentic equilibrium state, while the one
shown in Fig.~3(a) of Ref.~\cite{rotate two1} corresponds to a transition
state.

\begin{figure}[b]
\begin{center}
\vspace{-0.0cm} \includegraphics[width=8.5cm]{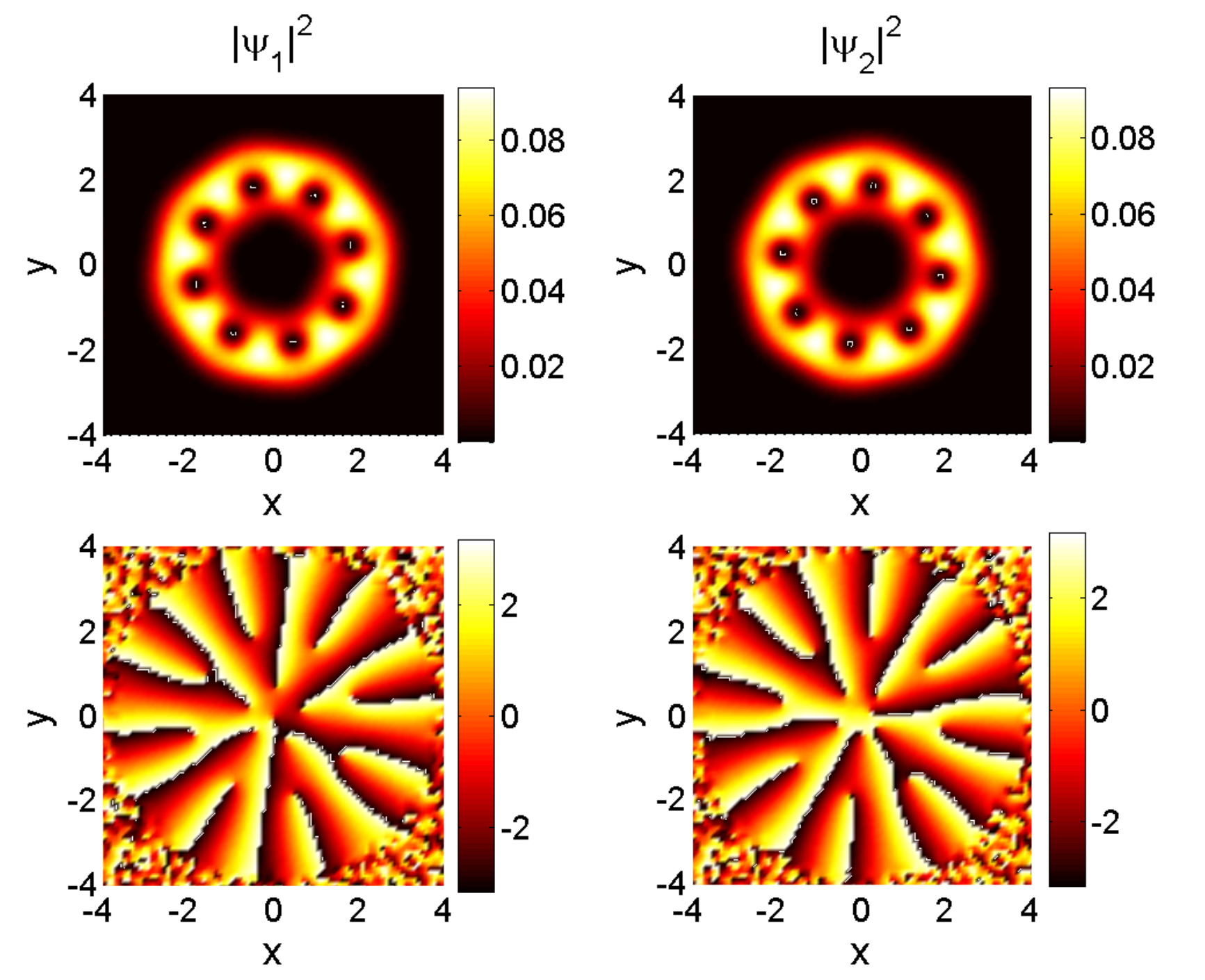}
\end{center}
\par
\vspace{-0.7cm}
\caption{(Color online) Vortex structures (top row) and phase profiles
(bottom row) of component 1 (left column) and 2 (right column) of
fast-rotating binary-mixture condensates. Here $g_{11}=55$, $\protect\alpha %
_{2}=1$, $\protect\alpha _{12}=0.5$, $\protect\lambda =1$, and $\Omega =2.5$%
. }
\label{fig4}
\end{figure}

Fig.~\ref{fig3} shows the vortex structures of the mixtures with $%
g_{11}=1300 $, $\alpha _{2}=0.5$, and $\alpha _{12}=0.8$. For these
parameters, the system is still in the VL regime with $\Omega <\Omega _{c}$.
In this case, the system has a ball-and-shell non-rotating ground state
(phase III) and in the vortex state it forms an interlocking oniony
vortex-sheet structure.

\subsubsection{VLH state with $\Omega>\Omega_{c}$}

The results in the VLH regime with $\Omega >\Omega _{c}$ are considered
next. In this subsection, to see more clearly the vortex physics, we show
both results of both density profile $n_{j}(x,y)=|\psi _{j}(x,y)|^{2}$ and
phase profile given by 
\begin{equation}
S_{j}(x,y)=\arctan \left[ {\frac{\mathrm{Im}\psi _{j}(x,y)}{\mathrm{Re}\psi
_{j}(x,y)}}\right] .  \label{phasej}
\end{equation}%
In the phase profile, the end point of the boundary between a $\pi $ phase
line and a $-\pi $ phase line will correspond to a vortex. In addition, the
circulation and the number of vortices can also be counted directly. Fig.~%
\ref{fig4} shows the vortex structure and the corresponding phase profile of
fast-rotating binary-mixture condensates with $g_{11}=55$, $\alpha _{2}=1$,
and $\alpha _{12}=0.5$. The ground state of the corresponding non-rotating
condensate mixture is miscible (phase I) to which the wavefunctions of the
two components overlap entirely. In view of Fig.~\ref{fig4}, it is found
that the two annular vortex arrays are interlocking in a manner that density
peak of one component is located at the density hole of the other component.

\begin{figure}[t]
\begin{center}
\vspace{-0.5cm} \includegraphics[width=8.5cm]{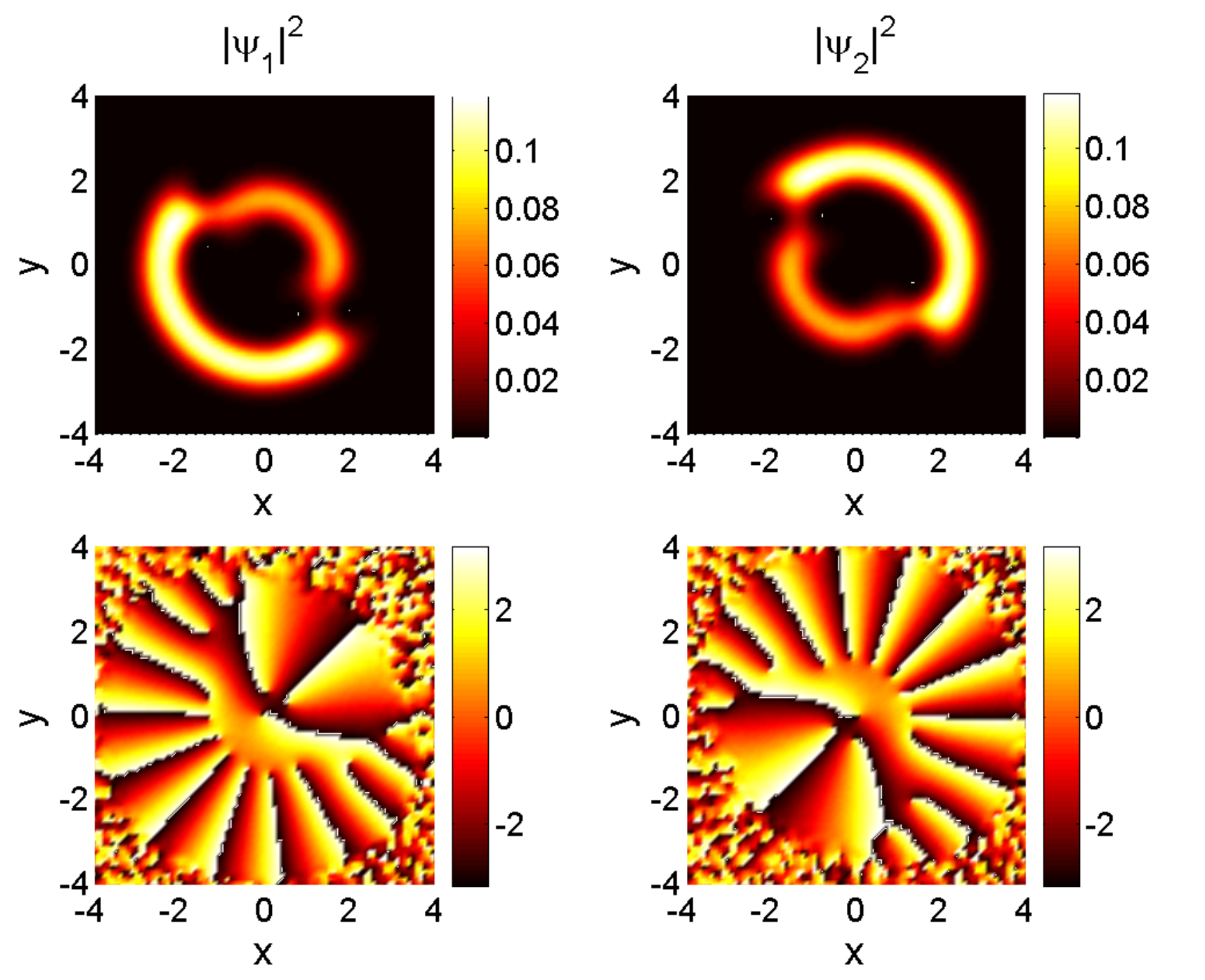}
\end{center}
\par
\vspace{-0.7cm}
\caption{(Color online) Vortex structures (top row) and phase profiles
(bottom row) of component 1 (left column) and 2 (right column) of
fast-rotating binary-mixture condensates. Here $g_{11}=55$, $\protect\alpha %
_{2}=1$, $\protect\alpha _{12}=1.3$, $\protect\lambda =1$, and $\Omega =2.5$%
. }
\label{fig5}
\end{figure}

Fig.~\ref{fig5} shows the vortex structure and the corresponding phase
profile of the mixture with $g_{11}=55$, $\alpha _{2}=1$, and $\alpha
_{12}=1.3$. The non-rotating counterpart has an asymmetric separated ground
state (phase II in Fig.~\ref{fig1}). Due to the strong repulsive interaction
between two condensates which results the asymmetric separated
characteristic, the two condensates occupy on the opposite side of each
other.

\begin{figure}[t]
\begin{center}
\vspace{-0.0cm} \includegraphics[width=8.5cm]{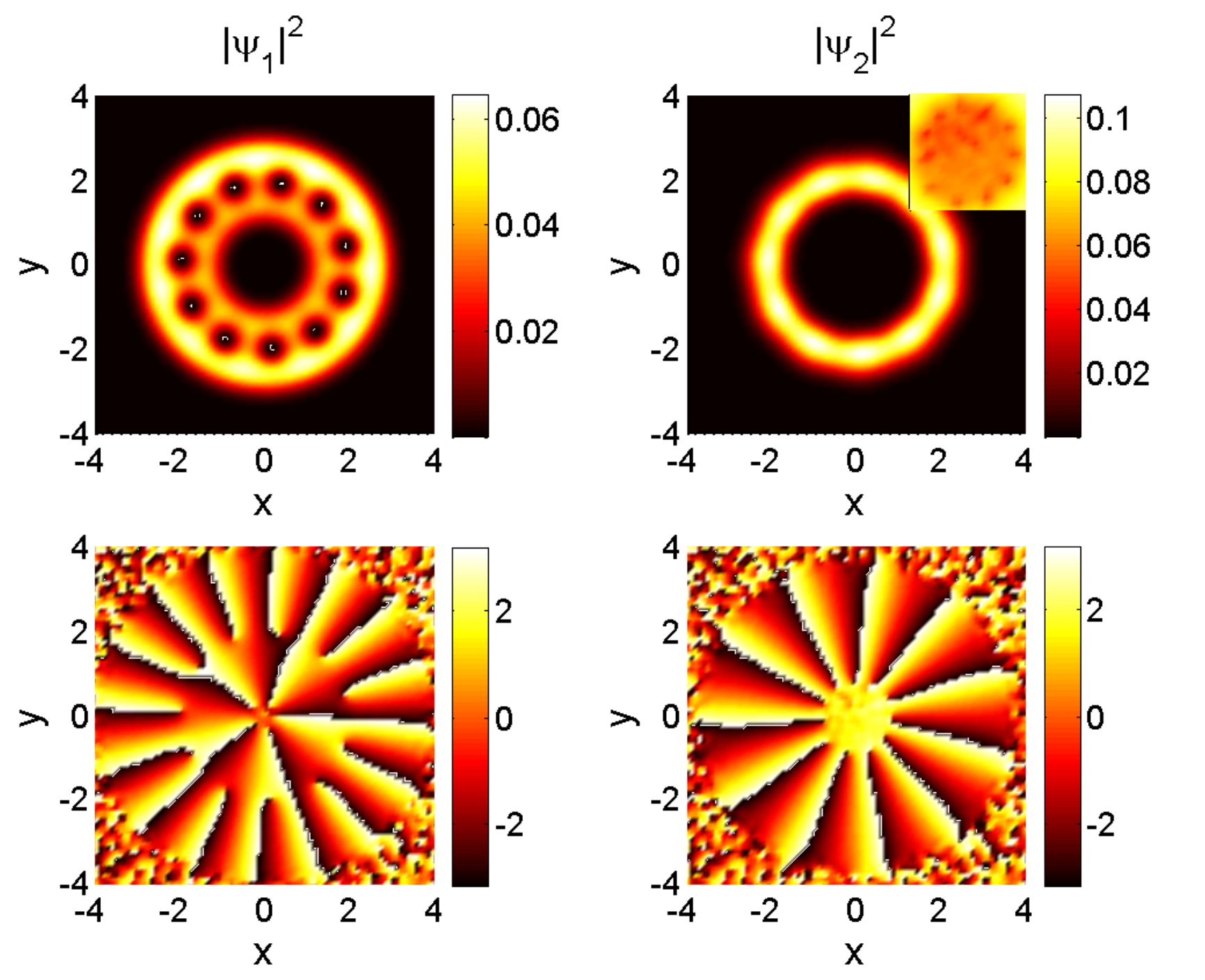}
\end{center}
\par
\vspace{-0.7cm}
\caption{(Color online) Vortex structures (top row) and phase profiles
(bottom row) of component 1 (left column) and 2 (right column) of
fast-rotating binary-mixture condensates. Here $g_{11}=55$, $\protect\alpha %
_{2}=0.5$, $\protect\alpha _{12}=0.8$, $\protect\lambda =1$, and $\Omega
=2.5 $. The inset shows a more clear view of the vortices in component 2.}
\label{fig6}
\end{figure}

Fig.~\ref{fig6} shows the vortex structure and the corresponding phase
profile of a condensate mixture with $g_{11}=55$, $\alpha _{2}=0.5$, $\alpha
_{12}=0.8$. With respect to phase III in Fig.~\ref{fig1}, the non-rotating
mixture has a ball-and-shell ground state, \textit{i.e.}, the component with
larger intraspecies interaction occupying the outside and forming a shell,
while the component with smaller intraspecies interaction occupying the
inside and forming a ball. It is found that vortices in component 1 (of
larger intraspecies interaction) form a circular array around the central
low-density hole, while vortices in component 2 (of smaller intraspecies
interaction) also form a circular array which is closer to the center (see
the inset of the density profile or the phase profile plot). In addition to
the vortices, the \textquotedblleft ball" of the non-rotating counterpart of
component 2 is actually pushed away (due to fast rotation) from the center
and forms a robust ringlike condensate located at where the vortices of
component 1 are (\emph{i.e.}, interlocking). This robustness actually
resists its own vortices revolving into it.

\begin{figure}[t]
\begin{center}
\vspace{-0.0cm} \includegraphics[width=8.5cm]{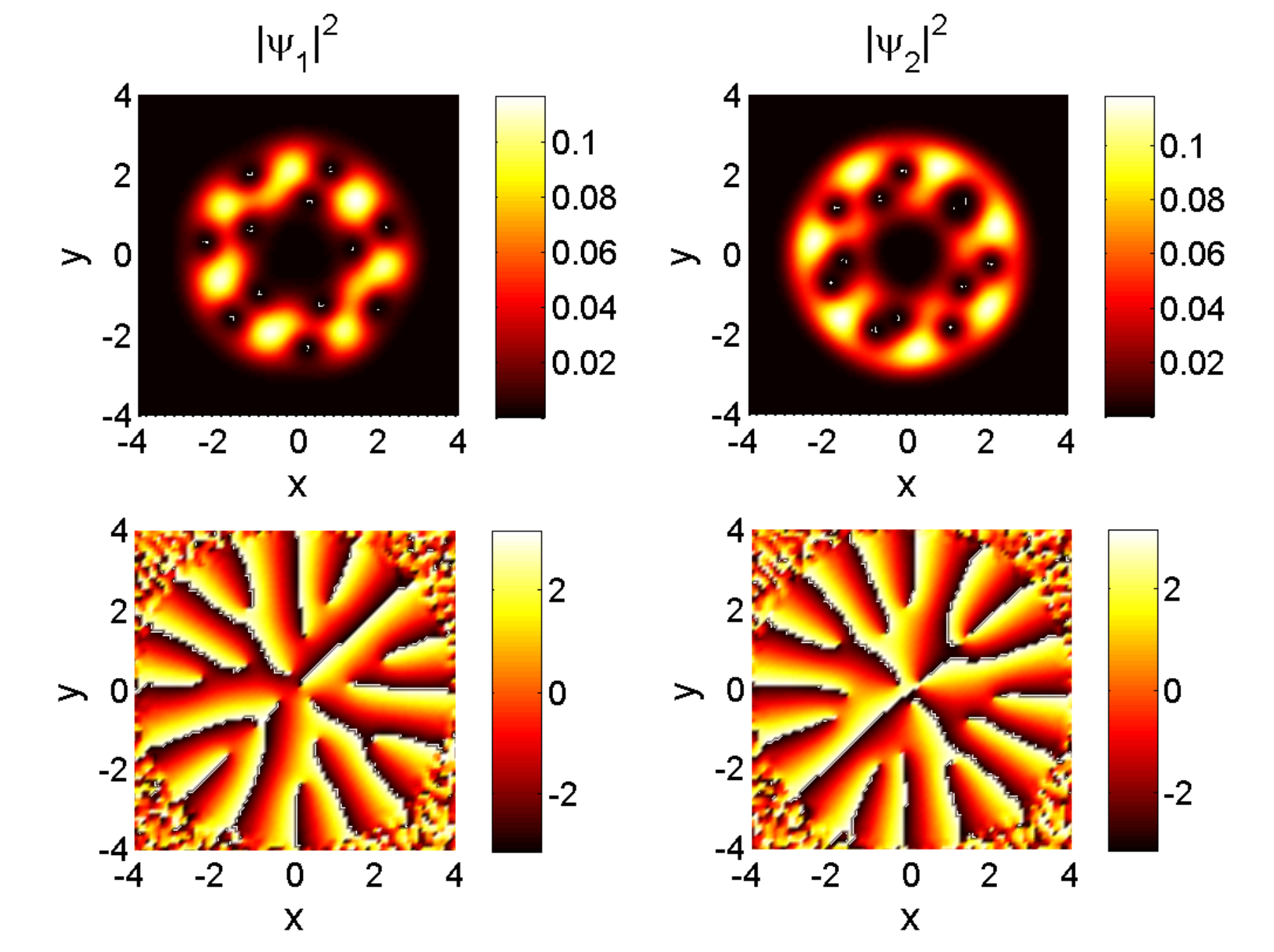}
\end{center}
\par
\vspace{-0.7cm}
\caption{(Color online) Vortex structures (top row) and phase profiles
(bottom row) of component 1 (left column) and 2 (right column) of
fast-rotating binary-mixture condensates. Here $g_{11}=55$, $\protect\alpha %
_{2}=1$, $\protect\alpha _{12}=1$, $\protect\lambda =1$, and $\Omega =2.5$.}
\label{fig7}
\end{figure}

In Fig.~\ref{fig7}, we have also shown the vortex structure and its
corresponding phase profile of a mixture at the isotropic point, with $%
g_{11}=55$, $\alpha _{2}=\alpha _{12}=1$. Similar to the result of the
middle row of Fig.~\ref{fig2}, fast-rotating condensates tend to form the
vortex-pair structure at the isotropic point.

\subsection{Concluding Remarks}

The overall features of vortex structures in ultrafast-rotating
binary-mixture condensates can be understood as follows. Pertaining to phase
I in Fig.~\ref{fig1}, when $\alpha_{12}=0$ where the two components are not
interacting with each other, the theory is essentially reduced to the one
for single components. In this limit, triangular vortex lattices are
expected to form with $\Omega<\Omega_{c}$, while annular vortex arrays are
expected to form with $\Omega>\Omega_{c}$. As $\alpha_{12}$ is present and
increases, vortex cores of one component gradually shift away from those of
the other component and consequently with $\Omega<\Omega_{c}$ the triangular
lattices are distorted. Eventually the vortices for each component will form
a square lattice instead of a triangular one. As $\alpha_{12}$ exceeds $%
\alpha_{2}$, equivalently for the system to shift to phase II or III, the
condensates can undergo phase separation to spontaneously form domains. For
one condensate, the cavity of another condensate is where the lower
effective potential is, which is more apt to be occupied. However, more
interlocking will cause wavefunctions overlap more and at the same time
raise the interspecies interaction energy $\propto g_{12}\left\vert
\Psi_{1}\right\vert ^{2}\left\vert \Psi_{2}\right\vert ^{2}$. In order to
prevent the above-mentioned interlocking that causes high energies, the
vortices are actually concentrated out of the condensates and form the
vortex sheets. Consequently the two components form complementary structures
to each other and the total density will be roughly described by the
Thomas-Fermi distribution $|\Psi_{1}|^{2}+|\Psi_{2}|^{2}\propto\max [\mu_{%
\mathrm{TF}}-(m\omega^{2}r^{2}/2+ur^{4}/4),0]$. The results in the bottom
row of Fig.~\ref{fig2} as well as in Figs.~3, 5, and 6 are examples of this
kind.

\section{Conclusion}

\label{sec:conclusion}

This paper investigates the equilibrium vortex structures of
ultrafast-rotating binary-mixture condensates trapped in a
harmonic-plus-quartic potential. In contrast to the harmonic trap alone case
where the system is unstable when the rotation frequency $\Omega_{0}$ is
higher than the radial trap oscillator frequency $\omega$, the added quartic
trap can lead the system to remain stable at higher rotation velocity ($%
\Omega_{0}>\omega$). Due to the complexity of interactions in the binary
system, there often occur many metastable states in the fast-rotating
two-component condensate system and the standard imaginary-time propagating
approach may not really converge to the true equilibrium states of the
system. In this regard, we have applied a combined numerical scheme to
effectively assure that the density profiles do really saturate at
sufficiently low temperatures. A critical rotating frequency $\Omega_{c}$
which characterizes the transition between the VL and the VLH states is
identified. Under high rotation frequencies ($\Omega_{0}>\omega$), a variety
of vortex structures of the two-component condensates are shown for $%
\Omega<\Omega_{c}$, similar to those presented in Ref.~\cite{rotate two1},
and also for $\Omega>\Omega_{c}$ in particular to which various annular
vortex structures occur.


\section{Acknowledgments}

This work was supported by National Science Council of Taiwan (Grant Nos.
99-2112-M-003-006-MY3 and 98-2112-M-018-001-MY2). We also acknowledge the
support from the National Center for Theoretical Sciences, Taiwan.


\begin{thebibliography}{99}
\bibitem{vortex1} M. R. Matthews, B. P. Anderson, P. C. Haljan, D. S. Hall,
C. E. Wieman, and E. A. Cornell, Phys. Rev. Lett. \textbf{83}, 2498 (1999).

\bibitem{lattice1} J. R. Abo-Shaeer, C. Raman, J. M. Vogels, and W.
Ketterle, Science \textbf{292}, 476 (2001).

\bibitem{lattice2} C. Raman, J. R. Abo-Shaeer, J. M. Vogels, K. Xu, and W.
Ketterle,Phys. Rev. Lett. \textbf{87}, 210402 (2001).

\bibitem{lattice3} P. C. Haljan, I. Coddington, P. Engles, and E. A.
Cornell, Phys. Rev. Lett. \textbf{87}, 210403 (2001).

\bibitem{lattice4} P. Engles, I. Coddington, P. C. Haljan, and E. A.
Cornell, Phys. Rev. Lett. \textbf{89}, 100403 (2002).

\bibitem{exp} C. J. Myatt, E. A. Burt, R. W. Ghrist, E. A. Cornell, and C.
E. Wieman, Phys. Rev. Lett. \textbf{78}, 586 (1997).

\bibitem{sym-asym1} B. D. Esry, and C. H. Greene, Phys. Rev. A \textbf{59},
1457 (1999).

\bibitem{sym-asym} M. Trippenbach, K. G\'{o}ral, K. Rz\c{a}\.{z}ewski, B.
Malomed, and Y B Band, J. Phys. B \textbf{33,} 4017 (2000).

\bibitem{sym-asym2} F. Riboli, and M. Modugno, Phys. Rev. A \textbf{65},
063614 (2002).

\bibitem{sym-asym3} D. M. Jezek, and P. Capuzzi, Phys. Rev. A \textbf{66},
015602 (2002).

\bibitem{sym-asym4} A. A. Svidzinsky, and S. T. Chui, Phys. Rev\textit{.} A 
\textbf{67}, 053608 (2003).

\bibitem{sym-asym5} C. C. Huang, and W. C. Wu, Phys. Rev. A \textbf{75},
023609 (2007).

\bibitem{rotate two} E. J. Mueller and T. L. Ho, Phys. Rev. Lett. 88, 180403
(2002).

\bibitem{rotate two1} K. Kasamatsu, M. Tsubota, and M. Ueda, Phys. Rev.
Lett. \textbf{91}, 150406 (2003).

\bibitem{rotate two2} S. J. Woo, S. Choi, L. O. Baksmaty, and N. P. Bigelow,
Phys. Rev A \textbf{75}, 031604(R) (2007).

\bibitem{multiply quantized1} E. Lundh, Phys. Rev. A, \textbf{65}, 043604
(2002).

\bibitem{multiply quantized2} K. Kasamatsu, M. Tsubota, and M. Usda, Phys.
Rev. A \textbf{66}, 053606 (2002).

\bibitem{multiply quantized3} P. Engels, I. Coddington, P. C. Haljan,V.
Schweikhard, and E. A. Cornell, Phys. Rev. Lett. \textbf{90}, 170405 (2003).

\bibitem{multiply quantized4} G. M. Kavoulakis and G. Baym, New J. Phys. 
\textbf{5}, 51 (2003).

\bibitem{multiply quantized5} V. Bretin, S. Stock, Y. Seurin, and J.
Dalibard, Phys. Rev. Lett. \textbf{92}, 050403 (2004).

\bibitem{multiply quantized6} T. P. Simula, A. A. Penckwitt, and R. J.
Ballagh, Phys. Rev. Lett. \textbf{92}, 060401 (2004).

\bibitem{multiply quantized7} A. L. Fetter, B. Jackson, and S. Stringari,
Phys. Rev. A \textbf{71}, 013605 (2005).

\bibitem{multiply quantized8} C. C. Huang, C. H. Liu, and W. C. Wu, Phys.
Rev. A \textbf{81}, 043605 (2010).

\bibitem{phenomenological damping} S. Choi, S. A. Morgan, and K. Burnett,
Phys. Rev. A \textbf{57}, 4057 (1997).

\bibitem{SGPE Stoof1} H. T. C. Stoof, J. Low Temp. Phys. \textbf{114}, 11
(1999).

\bibitem{SGPE Stoof2} H. T. C. Stoof, J. Low Temp. Phys. \textbf{124}, 431
(2001).

\bibitem{SPGPE Gardiner1} C. W. Gardiner, J. R. Anglin, and T. I. A. Fudge,
J. Phys. B \textbf{35}, 1555 (2002).

\bibitem{SPGPE Gardiner2} C. W. Gardiner and M. J. Davis, J. Phys. B \textbf{%
36}, 4731 (2003).

\bibitem{vortex SPGPE} A. S. Bradley, C. W. Gardiner, and M. J. Davis, Phys.
Rev. A \textbf{77}, 033616 (2008).

\bibitem{antiferromagnetism} As discussed in Ref.~\cite{rotate two1},
interaction energy $E_{int}$ can be expressed in term of the total density $%
n=\left\vert \Psi_{1}\right\vert ^{2}+\left\vert \Psi_{2}\right\vert ^{2}$
and the spin variable $S=\left\vert \Psi_{1}\right\vert ^{2}-\left\vert
\Psi_{2}\right\vert ^{2}$ as $E_{int}=\left( C_{11}/8\right) \int d\mathbf{\
r}[\left( 1+\alpha_{2}+2\alpha_{12}\right) n^{2}+\left(
1+\alpha_{2}-2\alpha_{12}\right) S^{2}+2\left( 1-\alpha_{2}\right) nS]$. If
the coefficient $\left( 1+\alpha_{2}-2\alpha_{12}\right) $ is positive,
antiferromagnetism manifests and makes a square lattice stabilized.
\end{thebibliography}
\end{document}